\documentclass[twocolumn,showpacs,pra]{revtex4} 

\usepackage{amssymb, amsfonts, amsmath}
\usepackage{graphicx, subfigure}
\usepackage{dcolumn}
\usepackage{bm}

\newcommand{\rd}{{\rm d}}
\newcommand{\re}{{\rm e}}
\newcommand{\ri}{{\rm i}}

\begin{document}

\title{AC-induced superfluidity}

\author{Andr\'e Eckardt}
\email{eckardt@theorie.physik.uni-oldenburg.de}
\author{Martin Holthaus}

\affiliation{                    
  Institut f\"ur Physik, Carl von Ossietzky Universit\"at, D-26111 Oldenburg,
  Germany}

\date{October 10, 2007}

\begin{abstract}
	{We argue that a system of ultracold bosonic atoms in a tilted
	optical lattice can become superfluid in response to resonant 
	AC forcing. Among others, this allows one to prepare a Bose--Einstein
	condensate in a state associated with a negative effective mass. 
	Our reasoning is backed by both exact numerical simulations for 
	systems consisting of few particles, and by a theoretical approach 
	based on Floquet--Fock states. }
\end{abstract}


\maketitle

\section{Introduction}
The study of ultracold atoms in optical lattices by now has opened up a 
promising new area of research on the borderlines between atomic and molecular 
physics, quantum optics, and condensed-matter physics: Many-body systems of 
paradigmatic importance to condensed-matter theory are being experimentally 
realized, with the precision and flexibility offered by quantum optical tools,
by means of interacting ultracold atoms subjected to periodic potentials 
generated by standing waves of laser light. A hallmark example along this
line is provided by the successful implementation of the 
Bose--Hubbard Hamiltonian~\cite{FisherEtAl89,JakschEtAl98}, and the 
observation of the predicted quantum phase transition from a superfluid
to a Mott insulator~\cite{GreinerEtAl02,StoeferleEtAl04,MunEtAl07}. 
Extrapolating these developments, it has been suggested to employ fermionic 
atoms confined by optical lattices in order to achieve a laboratory 
realization of the (fermionic) Hubbard model, and thus to tackle questions 
concerning high-$T_c$-superconductivity by performing measurements on that 
system~\cite{HofstetterEtAl02}, while an exact numerical treatment of the
very model lies still beyond existing computational resources.   

In this letter, we propose to investigate time-dependent many-body dynamics 
of ultracold bosonic atoms in a one-dimensional optical lattice subjected 
to both a static tilt and an additional time-periodic (AC) drive. We argue 
that such a system exhibits an effect which has no immediate solid-state 
analogue, namely, the appearance of a superfluid state in response to resonant 
forcing. We first discuss the underlying mechanism in a general manner, before 
presenting numerical calculations for small systems, and a quantum Floquet 
theoretical approach which both support our deductions. Finally we provide
parameter estimates concerning the experimental verification of this effect.

\section{The basic mechanism}
\begin{figure}
\centering
\includegraphics[width=0.6\linewidth]{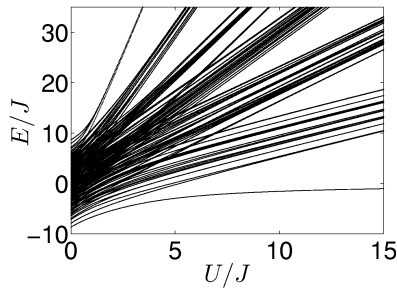}
\caption{\label{F_1}
	Energy spectrum versus interaction strength $U/J$ for a small,
	untilted and undriven Bose--Hubbard-system~(\ref{eq:BHS}) with
	$N = 5$ particles on $M = 5$ sites. The splitting-off of the
	ground state around $U/J \approx 4$ is a precursor of the quantum
	phase transition that occurs in an infinitely large system.}
\end{figure}

Ultracold bosonic atoms in a sufficiently deep one-dimensional optical 
lattice with $M$ sites are described, to good approximation, by the 
Bose--Hubbard model~\cite{FisherEtAl89,JakschEtAl98}
\begin{equation}
	\hat{H}_0 = -J \sum_{\ell = 1}^{M-1}\left( 
	\hat{b}^\dagger_\ell \hat{b}^{\phantom\dagger}_{\ell+1} +
	\hat{b}^\dagger_{\ell+1} \hat{b}^{\phantom\dagger}_\ell \right)
	+ \frac{U}{2}\sum_{\ell=1}^M 
	\hat{n}_\ell \left( \hat{n}_\ell - 1 \right) \; ,
\label{eq:BHS}	
\end{equation}
where $\hat{b}^\dagger_\ell$ ($\hat{b}^{\phantom\dagger}_\ell$) denotes 
the creation (annihilation) operator for an atom in the Wannier state 
located at the $\ell$th lattice site, 
$\hat{n}_\ell = \hat{b}^\dagger_\ell \hat{b}^{\phantom\dagger}_\ell$
is the number operator for that site, the hopping matrix element~$J > 0$
parametrises the strength of tunneling between adjacent sites, and $U > 0$ 
quantifies the repulsion energy of a pair of atoms occupying the same
site. For vanishing interaction strength, that is, for $U/J \to 0$, 
all $N$ atoms condense into the lowest Bloch state of the lattice for 
temperature $T \to 0$, giving rise to a superfluid ground state
$|{\rm SF} \rangle = \frac{1}{\sqrt{N!}}\left(\frac{1}{\sqrt{M}}
	\sum_{\ell=1}^M \hat{b}^\dagger_\ell \right)^N |0\rangle$, where
$|0\rangle$ denotes the vacuum state. On the other hand, for vanishing 
tunneling contact, $U/J \to \infty$, the ground state is the Mott insulating 
state $|{\rm MI} \rangle = \prod_{\ell=1}^M
	\frac{(\hat{b}^\dagger_\ell)^n}{\sqrt{n!}}|0\rangle$, assuming
integer filling $n = N/M$. In the limit of a chain consisting of infinitely 
many sites, $M \to \infty$, a sharp transition between the superfluid 
regime with (quasi) long-range order and the Mott-insulating regime occurs 
at a critical value $(U/J)_{\rm c}$, accompanied by the emergence of a 
finite gap between the energy of the ground state and those of the excited 
states~\cite{FisherEtAl89}. For a filling factor of one particle per site, 
$N/M = 1$, one finds $(U/J)_{\rm c} \approx 3.4$~\cite{KuehnerEtAl00}. 
Remarkably, even calculations performed for quite small systems show a 
precursor of this quantum phase transition: In Fig.~\ref{F_1} we display the 
energy eigenvalues for a system with $N = M = 5$, as functions of the scaled 
interaction strength $U/J$. There is, of course, no sharp transition in such 
a small system, but the gradual splitting-off of the ground-state energy is 
clearly discernible.

When the Bose--Hubbard system~(\ref{eq:BHS}) with $M \to \infty$ is subjected 
to both a static tilt amounting to an energy difference per site of $K_0$, 
and to an AC force of frequency~$\omega$ and amplitude~$K_{\omega}$, the total 
Hamiltonian becomes $\hat{H}(t) = \hat{H}_0 + \hat{H}_1(t)$, where
\begin{equation}
	\hat{H}_1(t) = \big( K_0 + K_\omega \cos(\omega t) \big) 
	\sum_{\ell = 1}^M \ell \hat{n}_{\ell} \; . 
\label{eq:HIT}
\end{equation}
Assuming first $K_\omega = 0$, the static force effectuates a splitting of 
the single-particle Bloch band into a sequence of single-particle states
with equidistant energies $E_{\ell} = K_0\ell$ forming the so-called 
Wannier--Stark ladder~\cite{FukuyamaEtAl73}. The corresponding single-particle 
orbitals are associated with creation operators 
$\hat{c}^\dagger_\ell = \sum_i {\rm J}_{i-\ell}(2J/K_0) \hat{b}^\dagger_i$,
where ${\rm J}_\alpha(z)$ is a Bessel function of order~$\alpha$. Since
$|{\rm J}_0(2J/K_0)|^2 = 1 - 2(J/K_0)^2 + \mathcal{O}[(J/K_0)^4]$ for 
$J/K_0 \ll 1$, one has $\hat{c}^\dagger_\ell \approx \hat{b}^\dagger_\ell$
when $K_0 \gg J$, so that the Wannier--Stark states coincide approximately
with the original Wannier states, indicating that a strong tilt effectively
destroys the tunneling contact between adjacent lattice sites. 

The principle we are exploiting in the present proposal relies on the fact 
that a tunneling contact disabled by a uniform tilt can be partially 
restored when the system is driven resonantly, {\em i.e.\/}, when the 
driving frequency $\omega$ in eq.~(\ref{eq:HIT}) is chosen such that
\begin{equation}
	K_0 \approx \nu \hbar\omega
\label{eq:RES}
\end{equation}
with integer $\nu$, so that the energy of $\nu$ quanta $\hbar\omega$ 
bridges the energy difference between adjacent rungs of the Wannier--Stark
ladder~\cite{Zak93}. In that case the {\em driven\/} Bose--Hubbard system
behaves approximately, in a time-averaged sense, like an {\em undriven\/} 
system with a modified hopping matrix element
\begin{equation}
	J_{\rm eff} = (-1)^\nu {\rm J}_\nu(K_\omega/\hbar\omega) \, J \; , 
\label{eq:JEF}
\end{equation}
provided the frequency is sufficiently high, so that both $\hbar\omega \gg U$ 
und $\hbar\omega \gg J$. While such a modification of the tunneling contact is 
reminiscent of the AC-Josephson-effect~\cite{BaronePaterno82,EckardtEtAl05a}, 
here the many-body interaction has a remarkable consequence: When 
$|U/J_{\rm eff}|$ becomes smaller than $(U/J)_c$, the system becomes 
superfluid despite the presence of the strong tilt. Observe that this 
scenario also includes pure AC-forcing with $K_0 = 0$ as a special case for 
$\nu = 0$. For this case, which has been considered theoretically 
before~\cite{EckardtEtAl05b,CreffieldMonteiro06}, the expected
$J_0$-modification of the hopping matrix element has been beautifully confirmed
in a recent experiment~\cite{LignierEtAl07}, thus demonstrating the feasibility 
of exposing a Bose--Einstein condensate in an optical lattice to strong 
AC forcing without destroying its phase coherence. 

In order to verify the existence of AC-induced superfluidity, we 
propose the following experimental protocol: {\em (i)\/}~Initially the
ultracold atoms are prepared in the superfluid ground state of a shallow
one-dimensional optical lattice. Then the interaction parameter $U/J$
is ramped up by increasing the depth of the lattice. At the critical 
parameter $(U/J)_{\rm c}$ the system enters the Mott-insulator regime,
and the particles localise at the lattice sites. For large values of  
$U/J$, the system approximately reaches the extreme Mott state
$|{\rm MI} \rangle$, again assuming integer filling $n = N/M$. 
{\em (ii)\/}~Next, a strong  static tilt is applied, e.g.\ by accelerating 
the lattice. Then $U/J$ is switched down again to a small value, such that 
the system would fall into the superfluid regime if there were no tilt. 
However, it actually remains in a state close to $|{\rm MI} \rangle$, which, 
in its turn, still remains an approximate stationary state of the Hamiltonian, 
due to the strong localisation of the Wannier-Stark states. 
{\em (iii)\/}~Finally the resonant AC force is turned on. In order to guide 
the system adiabatically into the AC-induced superfluid state corresponding 
to the ground state of the effective undriven Bose--Hubbard Hamiltonian with 
the modified hopping element~(\ref{eq:JEF}), that force should be turned on 
smoothly, starting from $K_\omega/\hbar\omega = 0$.

\section{Exact simulation of small systems}
\begin{figure*}
\centering
\begin{minipage}{1\linewidth}
	\begin{minipage}{0.92\linewidth}
	\subfigure[ $\nu=0$]{\label{F_2a}
	\includegraphics[width = 0.32\linewidth]{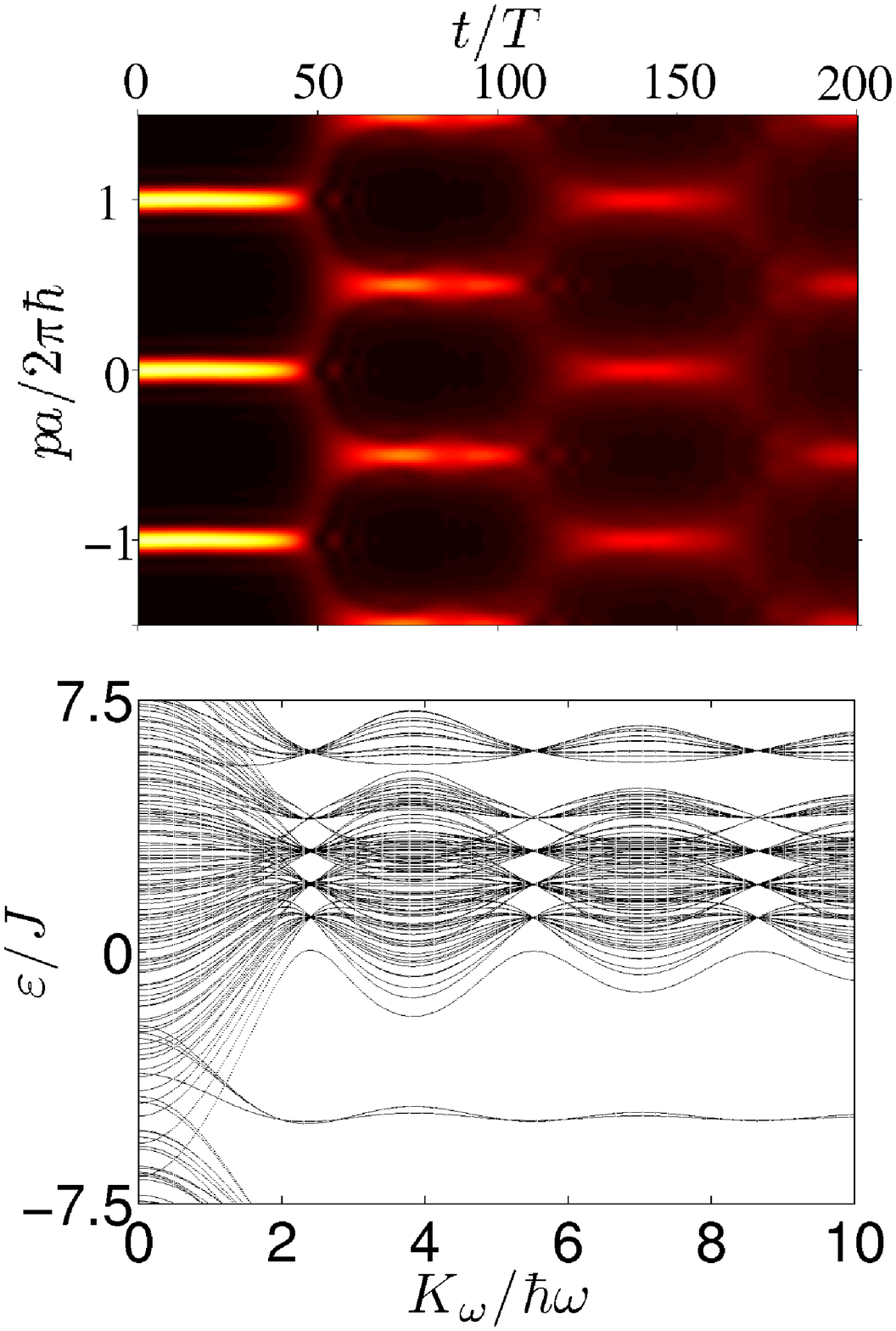}}
	\subfigure[ $\nu=1$]{\label{F_2b}
	\includegraphics[width = 0.32\linewidth]{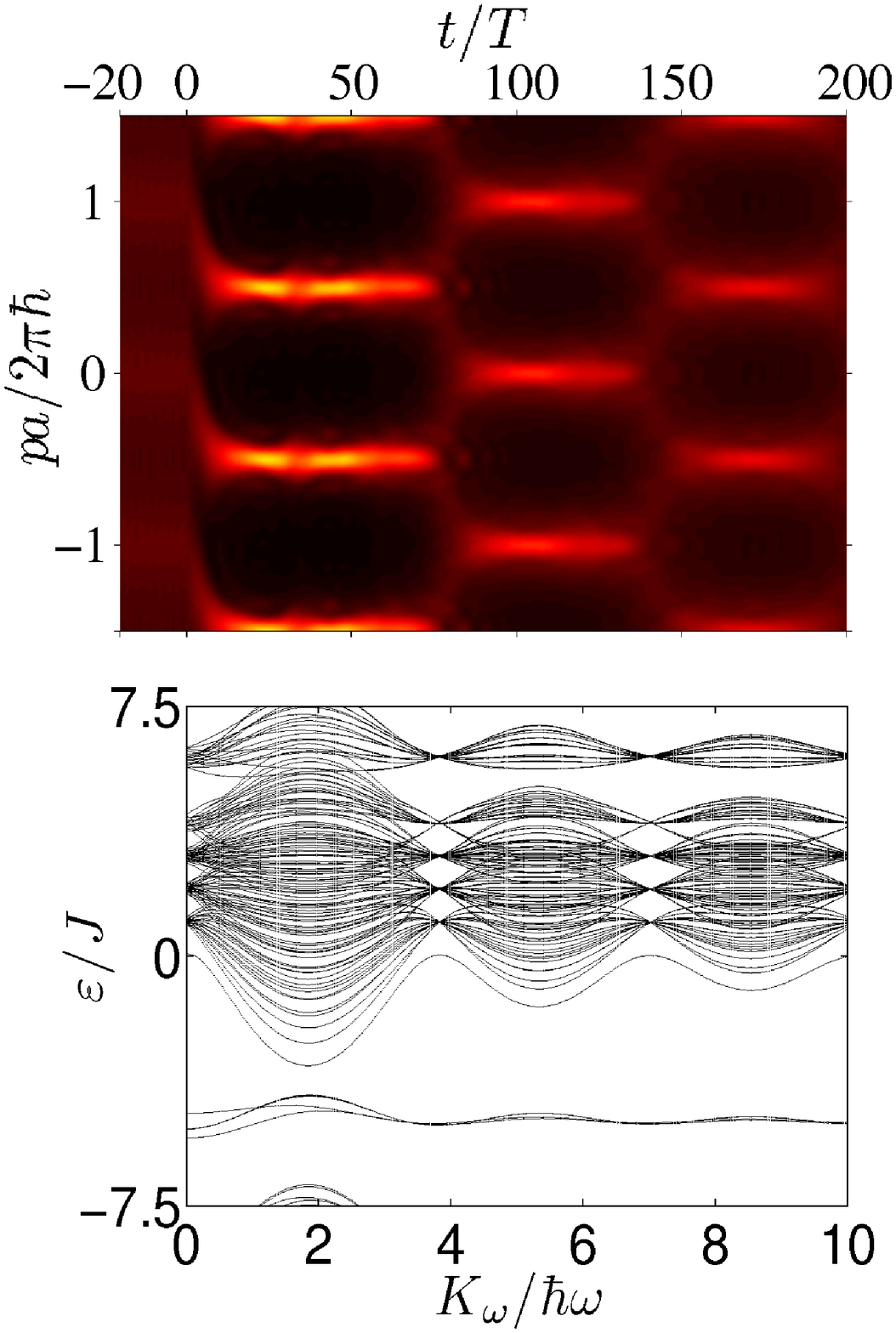}}
	\subfigure[ $\nu=2$]{\label{F_2c}
	\includegraphics[width = 0.32\linewidth]{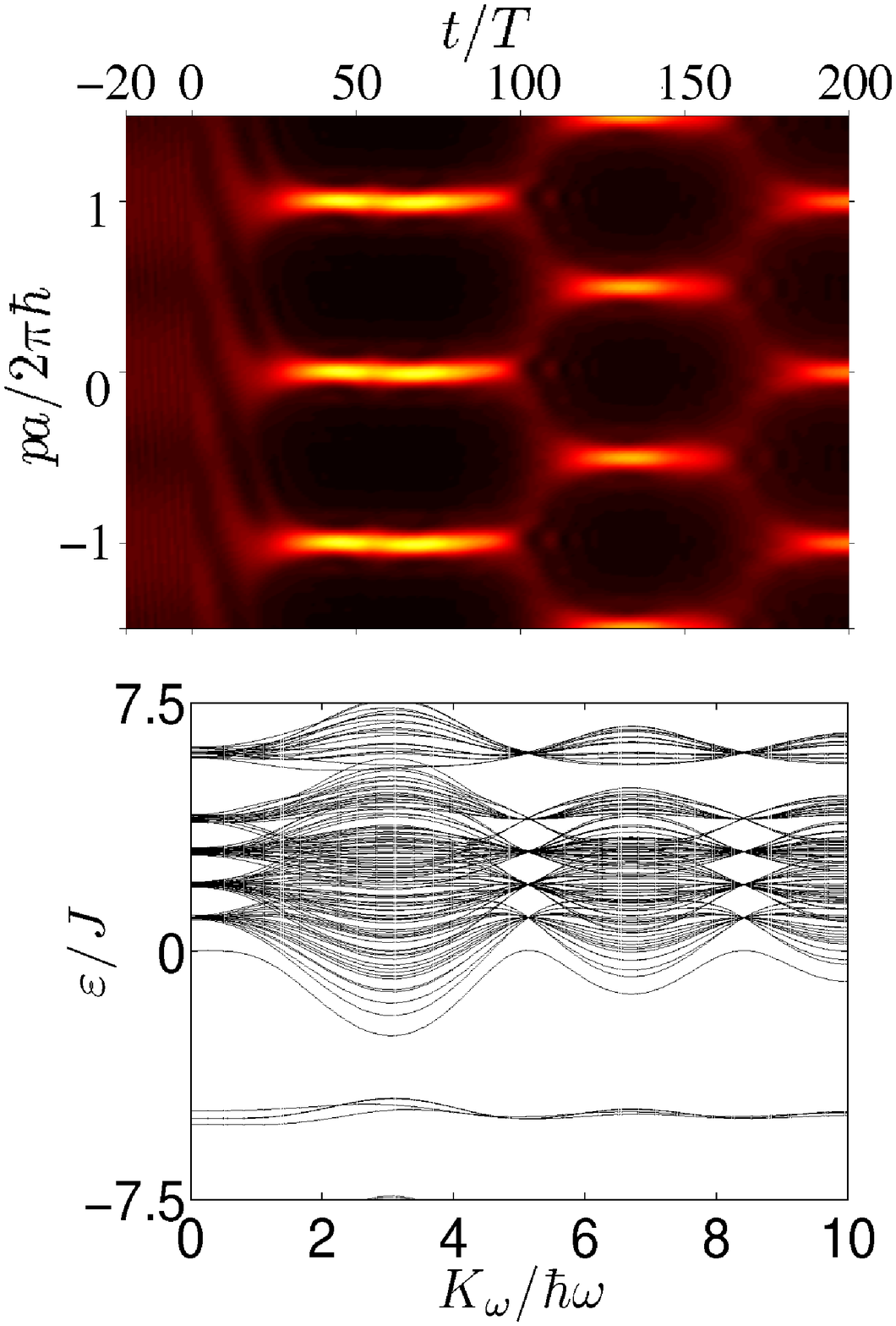}}
	\end{minipage}
	\begin{minipage}{0.04\linewidth}
	\includegraphics[width = 1\linewidth]{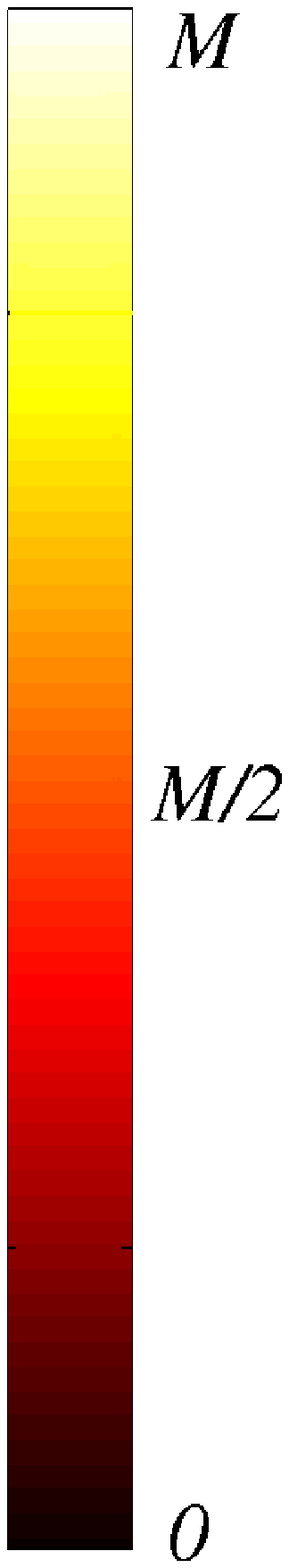}
	\end{minipage}
\end{minipage}
\caption{\label{F_2}
	{\em Upper row:\/} Three Brillouin zones of the quasimomentum
	distribution $\rho(p,t)$, recorded at integer $t/T$, while the driving
	amplitude $K_\omega/\hbar\omega$ increases linearly from $0$ to $10$
	between $t = 0\,T$ and $t = 200\,T$. Parameters for $t \geq 0\,T$
	are $U/J = 1$ and $\hbar\omega/J = 15$; $N = M = 9$.
	For $\nu=0$, the system starts in a superfluid state, corresponding
	to a sharply peaked distribution. When $K_\omega/\hbar\omega$ passes 
	a zero of ${\rm J}_0$, the maxima of the distribution switch from 
	the centers to the edges of the Brillouin zones, or vice versa.
	For $\nu = 1$ and $\nu = 2$, the system initially is prepared in
	a Mott-like state. Nonetheless, the sharp peaks emerging at later
	times, when $|J_{\rm eff}|$ is sufficiently large, signal the 
	appearance of AC-induced superfluidity. Again, the peaks switch 
	between the zone centers and the edges at the zeros of the Bessel 
	function ${\rm J}_1$ or ${\rm J}_2$, respectively.
	--- {\em Lower row:\/} Exact instantaneous quasienergies for the 
	same situations, computed with $N = M = 5$. Observe that the changes
	in the momentum distributions above are precisely reflected by
	the band collapses.}
\end{figure*}

We have computed the exact time evolution of a small system of $N = 9$
particles on $M = 9$ lattice sites, yielding the filling factor $N/M = 1$.
To analyze the time-dependent wave function $|\psi(t)\rangle$, we calculate 
the single-particle quasimomentum distribution 
\begin{equation}
	\rho(p,t) = \frac{1}{M}\sum_{\ell,j} 
	\exp\!\left[ \ri \frac{(\ell-j)p}{\hbar/a}\right]
	\langle \psi(t) | 
	\hat{b}^\dagger_\ell \hat{b}^{\phantom\dagger}_j
	| \psi(t) \rangle \; ,
\label{eq:QMD}
\end{equation}
where $a$ is the lattice constant, and record it versus $pa/(2\pi\hbar)$ at 
times~$t$ that are integer multiples of the driving period $T = 2\pi/\omega$.
This distribution is $2\pi\hbar/a$-periodic with respect to~$p$; together 
with an envelope provided by the momentum distribution of the corresponding 
Wannier state it yields the momentum distribution of the system that can be 
measured experimentally by time-of-flight absorption 
imaging~\cite{GreinerEtAl02,StoeferleEtAl04,MunEtAl07}. It allows one to 
clearly distinguish between the superfluid and the Mott state: Whereas the 
momentum distribution of a Mott state is rather structureless, a superfluid 
state with (quasi) long-range phase coherence is characterised by sharp peaks 
at the minima of the effective single-particle dispersion relation 
$E_{\rm eff}(p) = -2J_{\rm eff}\cos(pa/\hbar)$. 

Fixing the time scale such that the AC force with high frequency
$\hbar\omega/J = 15$ is turned on at $t = 0$, the system is initialised in 
the ground state of an undriven, untilted lattice with large interaction 
parameter $U/J = 50$ at time $t = -20\,T$. Simultaneously the lattice is 
tilted abruptly by $K_0 = \nu\hbar\omega$, with $\nu = 1$ or~$2$. At 
$t = -10\,T$ the interaction strength is decreased abruptly to $U/J = 1$. 
These preparatory steps lead to no significant change of the initial Mott 
state. That state coincides to good accuracy (overlap $\approx 0.95$ or 
larger) with the ground state $|{\rm MI}\rangle$ of the effective system 
for $K_\omega/\hbar\omega = 0$, i.e.\/, for $J_{\rm eff} = 0$. For 
comparison, we also show results obtained for an untilted lattice ($\nu = 0$). 
In that case, the system simply is initialised at $t = 0$ in its superfluid 
ground state for $U/J = 1$.

During the time intervall from $0\,T$ to $200\,T$ the dimensionless driving 
amplitude $K_\omega/\hbar\omega \equiv z$ is ramped up linearly from $0$ to 
$10$. The upper row in Fig.~\ref{F_2} shows the resulting quasimomentum 
distributions~(\ref{eq:QMD}), as obtained from numerical solutions of the 
time-dependent many-body Schr\"odinger equation. For $\nu = 0$, the system 
starts in a superfluid state. Accordingly, $\rho(p,t)$ displayed in 
Fig.~\ref{F_2a} shows a high contrast right from the outset at $t = 0$, with 
sharp maxima at the Brillouin zone center, 
$pa/(2\pi\hbar) = 0,\pm 1,\pm2,\ldots\;$.
However, when the gradually rising amplitude $K_\omega(t)/\hbar\omega$ 
passes the first zero of ${\rm J}_0(z)$ at $z \approx 2.4$, there is a short 
interval during which $|J_{\rm eff}|$ is so small, or $|U/J_{\rm eff}|$ so 
large, that the system becomes Mott-like~\cite{EckardtEtAl05b}. When the 
superfluid-like state reemerges at later $t$, the maxima of $\rho(p,t)$ have 
shifted to the edges of the Brillouin zone. This finding is fully in 
accordance with the fact that $J_{\rm eff}$ is {\em negative\/} between the 
first and second zero of ${\rm J}_0$, so that the minima of $E_{\rm eff}(p)$
occur at the zone edges. When the driving amplitude rises even beyond the
second zero of ${\rm J}_0(z)$ at $z \approx 5.5$, the original conditions
are restored. We remark that this change of sign of the effective hopping
matrix element, which has been clearly observed in the 
experiment~\cite{LignierEtAl07}, is a nontrivial consequence of the 
``dressing'' of the system achieved through the time-periodic forcing: In 
effect, this amounts to placing the condensate in a state with a negative 
effective mass, the sign of that mass being controlled by the forcing
strength.

\begin{figure}
\centering
\includegraphics[width=0.7\linewidth]{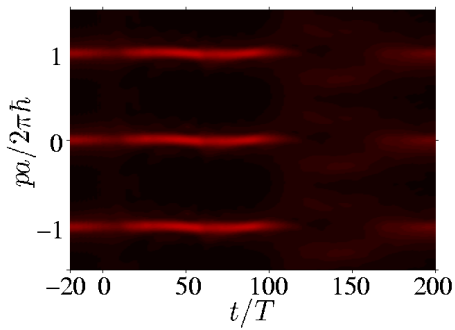}
\caption{\label{F_3}
	Momentum distribution for parameters and protocol as in 
	Fig.~\ref{F_2c}, but for an initial state with half filling,
	$N = 6$~particles on $M = 12$~sites. Shortly after $t = 100\,T$,
	when $J_{\rm eff}$ changes its sign, the signature of superfluidity
	is lost, since the system is unable to follow the rapidly changing
	ground state.}
\end{figure}

When $\nu = 1$, the system initially is Mott-like at $t = 0\,T$. However, as 
soon as $K_\omega(t)/\hbar\omega$ becomes sufficently large, the tunneling 
contact is restored, and the system beomes superfluid-like, as wittnessed 
by the sharp feature visible in Fig.~\ref{F_2b}. This is precisely the effect 
we are focussing on. Observe that here $J_{\rm eff}$ is negative between 
$z = 0$ and the first zero of ${\rm J}_1(z)$ at $z \approx 3.8$, so that the 
negative-mass picture now holds even for small driving amplitudes. The results 
for $\nu = 2$ in Fig.~\ref{F_2c} likewise follow the expected pattern. 
Thus, the computations summarised in the upper row of Fig.~\ref{F_2} can 
be explained in terms of an effective time-independent Bose--Hubbard 
Hamiltonian~(\ref{eq:BHS}) equipped with the modified hopping matrix 
element~(\ref{eq:JEF}) at any instantaneous value of the driving amplitude, 
in combination with approximate adiabatic following of the system's wave 
function in response to the actual slow rise of that amplitude. This 
intuitive picture has a striking implication: A Bose--Hubbard system at half 
filling remains superfluid even for $J/U \approx 0$~\cite{FisherEtAl89}. 
But when the sign of $J_{\rm eff}$ changes, the superfluid ground state 
changes abruptly, rendering adiabatic following impossible. This is illustrated
in Fig.~\ref{F_3}, where we have plotted the momentum distribution for the 
same parameters and protocol as previously considered in Fig.~\ref{F_2c}, 
but for $N = 6$ and $M = 12$: For times slightly larger than $100\,T$, where 
$J_{\rm eff}$ becomes negative, superfluidity is lost. This expected outcome 
signals that the high degree of adiabaticity achieved in Fig.~\ref{F_2} for 
unit filling actually is a nontrivial many-body effect; it hinges on the 
presence of a Mott-like state when $J_{\rm eff}/U$ is close to zero.

\section{Quantum Floquet theoretical approach}
We now give a formal explanation of the approximate AC-induced modification 
of the hopping matrix element~(\ref{eq:JEF}). We separate the full Hamiltonian
$\hat{H}_0 + \hat{H}_1(t)$ into a part which is site-diagonal,    
\begin{equation}
	\hat{H}_{\rm int}(t) = \frac{U}{2} 
	\sum_{\ell=1}^M  \hat{n}_\ell \left( \hat{n}_\ell - 1 \right) 
        + \hat{H}_1(t) \; ,
\end{equation}
and a part which describes tunneling between neighbouring sites,
\begin{equation} 
	\hat{H}_{\rm tun} = -J \sum_{\ell = 1}^{M-1}\left( 
	\hat{b}^\dagger_\ell \hat{b}^{\phantom\dagger}_{\ell+1} +
	\hat{b}^\dagger_{\ell+1} \hat{b}^{\phantom\dagger}_\ell \right) \; .
\label{eq:HTN}
\end{equation}
We do not require that the resonance condition~(\ref{eq:RES}) is met exactly, 
which would be close to impossible in a laboratory experiment, but rather 
admit a slight detuning, and choose the integer $\nu$ such that 
$|K_0 - \nu\hbar\omega|$ is minimised. Then one has 
\begin{equation}
	K_0 = \nu\hbar\omega + \Delta K_0
\end{equation}
with $|\Delta K_0| \ll \hbar\omega$. Now the set of Fock states
\begin{equation}
	| \{ n_\ell \} \rangle = \sum_\ell 
	\frac{(\hat{b}^\dagger_\ell)^{n_\ell}}{\sqrt{n_\ell!}} | 0 \rangle \; ,
\label{eq:SFB}
\end{equation}
where $\{ n_\ell \}$ runs over all admissible sets of occupation numbers 
of the lattice sites, forms a basis for the diagonalisation of the 
time-independent Bose--Hubbard Hamiltonian~(\ref{eq:BHS}). However, when the 
forcing~(\ref{eq:HIT}) is turned on with constant amplitude~$K_\omega$, the 
full Hamiltonian $\hat{H}(t) = \hat{H}_{\rm int}(t) + \hat{H}_{\rm tun}$  
is $T$-periodic in time, with $T = 2\pi/\omega$. In that case the system's 
Floquet states, {\em i.e.\/}, the complete set of explicitly time-dependent, 
$T$-periodic solutions $|u(t)\rangle\rangle$ to the eigenvalue equation
\begin{equation}
	\left[ \hat{H}(t) - \ri\hbar\partial_t \right] |u(t)\rangle\rangle 
	= \varepsilon \, |u(t)\rangle\rangle \; ,
\label{eq:EIG}
\end{equation}
take over the role of the stationary states; the wave functions
$|\psi(t)\rangle = |u(t)\rangle\rangle\exp(-\ri\varepsilon t/\hbar)$
solve the time-dependent Schr\"odinger equation. The spectrum of their 
quasienergies~$\varepsilon$ is obtained by diagonalising the operator
$\hat{H}(t) - \ri\hbar\partial_t$ in an extended Hilbert space of
$T$-periodic functions, endowed with the scalar product
\begin{equation}
	\langle\langle \, \cdot \, | \, \cdot \, \rangle\rangle
	\equiv \frac{1}{T} \int_0^T \! \rd t \,
	\langle \, \cdot \, | \, \cdot \, \rangle 
\label{eq:SCP}
\end{equation} 
which combines the usual scalar product 
$\langle \, \cdot \, | \, \cdot \, \rangle$ with time-avaraging~\cite{Sambe73}. 
Accordingly, we employ a set of Floquet--Fock states   
\begin{align}
&	| \{ n_\ell \},m\rangle\rangle \equiv | \{ n_\ell \} \rangle 
\nonumber \\ 
&	\; \times 
	\exp\!\left\{-\ri \left[ \frac{K_\omega}{\hbar\omega}\sin(\omega t)
	+ \nu\omega t \right] \sum_\ell \ell n_\ell + \ri m \omega  t \right\}
	& & 
\label{eq:FFB}
\end{align}
which diagonalise the part $\hat{H}_{\rm int}(t) - \ri\hbar\partial_t$ of the
quasienergy operator: By construction, one has 
\begin{align}
& 	\langle \langle \{ n_\ell \} , m | 
	\, \hat{H}_{\rm int}(t) - \ri\hbar\partial_t \, 
	| \{ n_\ell \} , m \rangle\rangle
\nonumber \\	
&	= 
	\frac{U}{2}\sum_\ell n_\ell \left( n_\ell - 1 \right)
	+ \Delta K_0 \sum_\ell \ell n_\ell + m \hbar\omega \; . 
\end{align}
Observe that the quasienergy spectrum is periodic in $\varepsilon$ with 
period~$\hbar\omega$; the integer $m = 0$, $\pm 1$, $\pm 2$, \ldots introduced 
as a Fourier index in the basis states~(\ref{eq:FFB}) serves to distinguish 
different Brillouin zones of that spectrum. Employing the identity
$
	\re^{\ri z \sin\varphi} = \sum_{k=-\infty}^{+\infty} 
	\re^{\ri k \varphi} {\rm J}_k(z) \; ,
$
one easily obtains 
\begin{align}
& 	\langle \langle \{ n_\ell' \} , m' | 
	\, \hat{b}^\dagger_\ell \hat{b}^{\phantom\dagger}_{\ell+1} \,
	| \{ n_\ell \} , m \rangle\rangle
\nonumber \\	
&	= \langle \{ n_\ell' \} | 
	\, \hat{b}^\dagger_\ell \hat{b}^{\phantom\dagger}_{\ell+1} \,
	| \{ n_\ell \} \rangle \,	
	{\rm J}_{s(m-m') - \nu}(K_\omega/\hbar\omega)
\end{align}
with $s \equiv \sum_\ell \ell (n_\ell - n_\ell') = +1$, since 
$\hat{b}^\dagger_\ell \hat{b}^{\phantom\dagger}_{\ell+1}$ transfers 
one particle by one site to the left. Note that when taking the adjoint 
matrix element, both $m$, $m'$ and $\{ n_\ell \}$, $\{ n_\ell' \}$
interchange their roles, so that the sign~$s$ changes to $-1$ and the 
numerical value of the element remains unchanged. Hence, with respect to 
the index~$m$ the matrix of the quasienergy operator has a simple block
structure:    
\begin{align}
& 	\langle \langle \{ n_\ell' \} , m' | 
	\, \hat{H}(t) - \ri\hbar\partial_t \, 
	| \{ n_\ell \} , m \rangle\rangle
\nonumber \\	
& 	= \delta_{m',m} \, \langle \{ n_\ell' \} | 
	\, \hat{H}_{\rm eff} + m\hbar\omega \, | \{ n_\ell \} \rangle 
\nonumber \\
&	\quad + (1 - \delta_{m',m}) \,
	\langle \{ n_\ell' \} | \, \hat{V} \, | \{ n_\ell \} \rangle \; . 
\label{eq:QEM}
\end{align}
The diagonal blocks ($m = m'$) are just the matrices of a Bose--Hubbard
system in the standard Fock basis~(\ref{eq:SFB}), without AC forcing,
but with modified hopping strength~(\ref{eq:JEF}) and possibly a slight 
residual tilt, as described by the effective Hamiltonian
\begin{align}
	\hat{H}_{\rm eff} \equiv & 
	-J_{\rm eff} \sum_{\ell = 1}^{M-1}\left( 
	\hat{b}^\dagger_\ell \hat{b}^{\phantom\dagger}_{\ell+1} +
	\hat{b}^\dagger_{\ell+1} \hat{b}^{\phantom\dagger}_\ell \right)
        + \frac{U}{2} 
	\sum_{\ell=1}^M  \hat{n}_\ell \left( \hat{n}_\ell - 1 \right)
\nonumber \\
	& + \Delta K_0 \sum_\ell \ell \hat{n}_\ell \; .
\label{eq:HEF}
\end{align}
These blocks are shifted against each other by multiples of the ``photon'' 
energy $\hbar\omega$, as corresponding to the Brillouin-zone structure of the 
quasienergy spectrum. They are coupled by off-diagonal blocks provided by
an operator $\hat{V}$ which differs from the tunneling term~(\ref{eq:HTN})
such that the hopping element $J$ is multiplied by a Bessel function  
${\rm J}_{m\!-\!m'\!-\!\nu}(K_\omega/\hbar\omega)$ 
for each particle transfer directed to the left, as described by a combination
$\hat{b}^\dagger_\ell \hat{b}^{\phantom\dagger}_{\ell+1}$,
and by ${\rm J}_{m'\!-\!m\!-\!\nu}(K_\omega/\hbar\omega)$
for each transfer directed to the right, corresponding to
$\hat{b}^\dagger_{\ell+1} \hat{b}^{\phantom\dagger}_\ell$.

The viewpoint advocated before, namely, the description of the forced system 
$\hat{H}_0 + \hat{H}_1(t) = \hat{H}_{\rm int}(t) + \hat{H}_{\rm tun}$
in terms of the effective time-independent Hamiltonian~(\ref{eq:HEF}), now
amounts to neglecting all off-diagonal coupling blocks with $m' \neq m$.
This is a viable approximation in the high-frequency regime, were 
(besides $|\Delta K_0|$) both relevant energy scales $J$ and $U$ are small 
compared to $\hbar\omega$. Intuitively speaking, this condition guarantees 
that the AC drive is off-resonant with respect to low-order transitions. 
As expressed by the scalar product~(\ref{eq:SCP}), the replacement of the 
original Hamiltonian by $\hat{H}_{\rm eff}$ essentially relies on the 
averaging principle.

\begin{figure*}
\centering
\subfigure[ $\nu=0$ and $K_\omega/\hbar\omega=1.811$]{\label{F_4a}
\includegraphics[width = 0.32\linewidth]{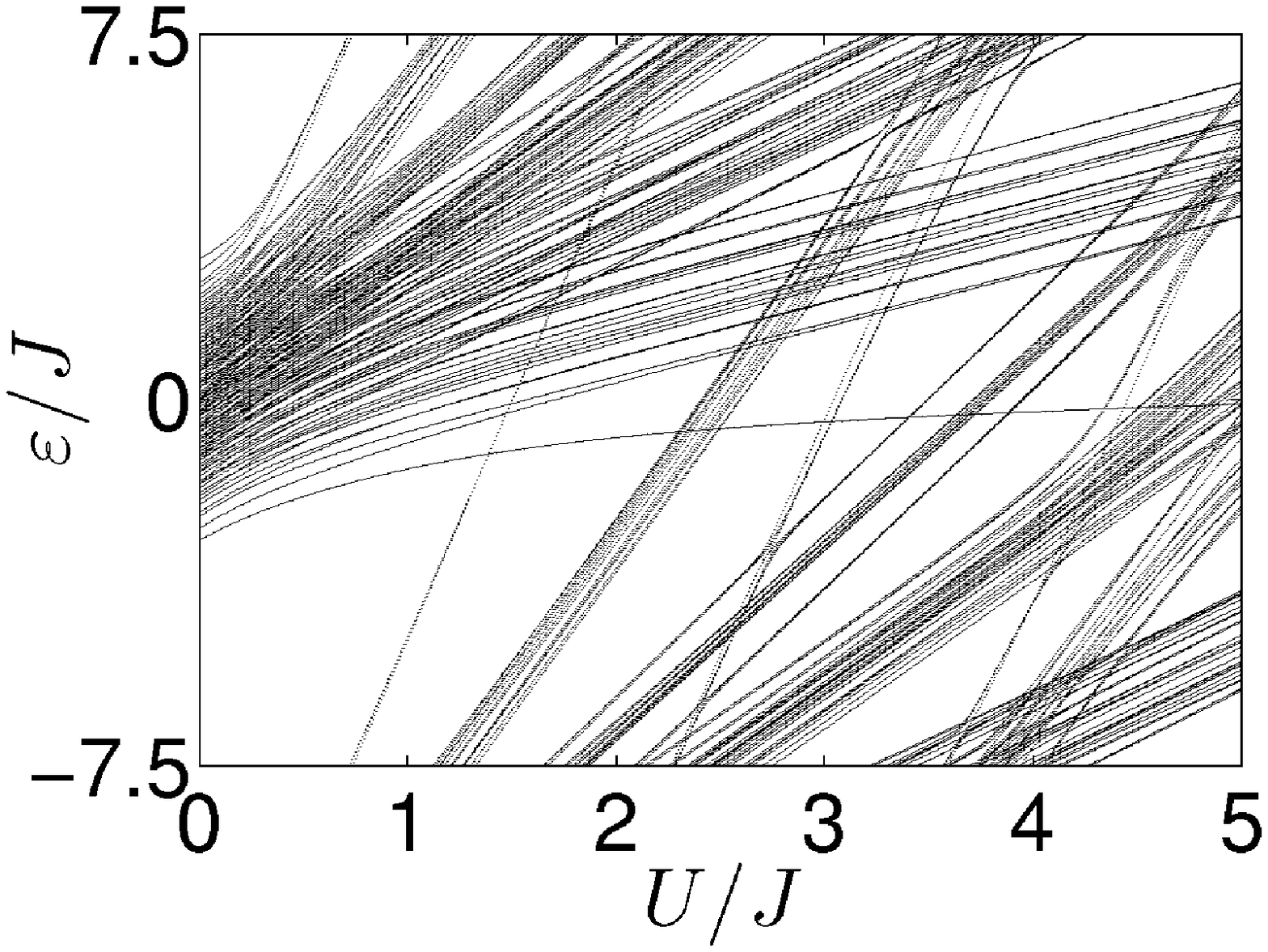}}
\subfigure[ $\nu=1$ and $K_\omega/\hbar\omega=0.711$]{\label{F_4b}
\includegraphics[width = 0.32\linewidth]{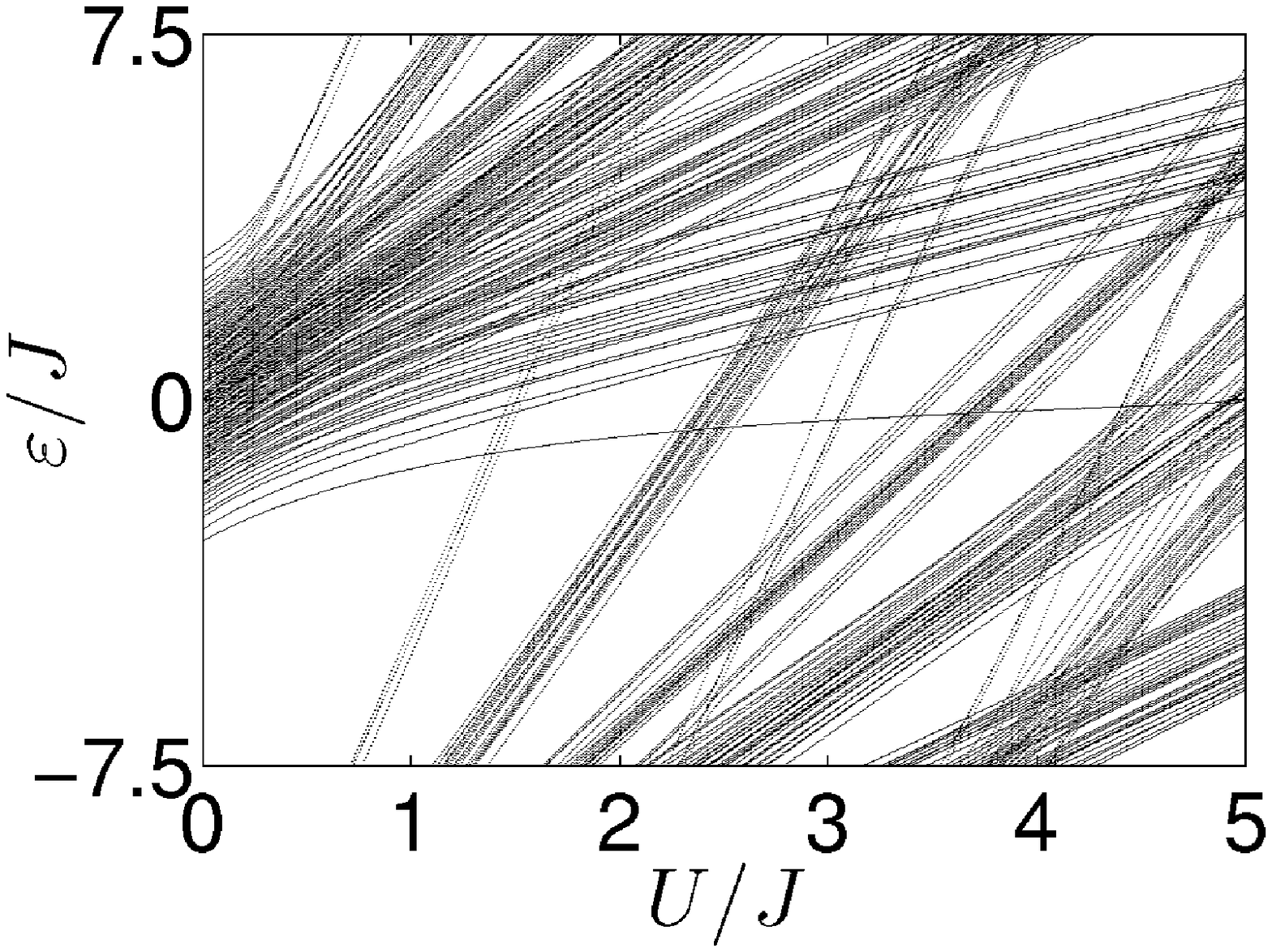}}
\subfigure[ $\nu=2$ and $K_\omega/\hbar\omega=1.915$]{\label{F_4c}
\includegraphics[width = 0.32\linewidth]{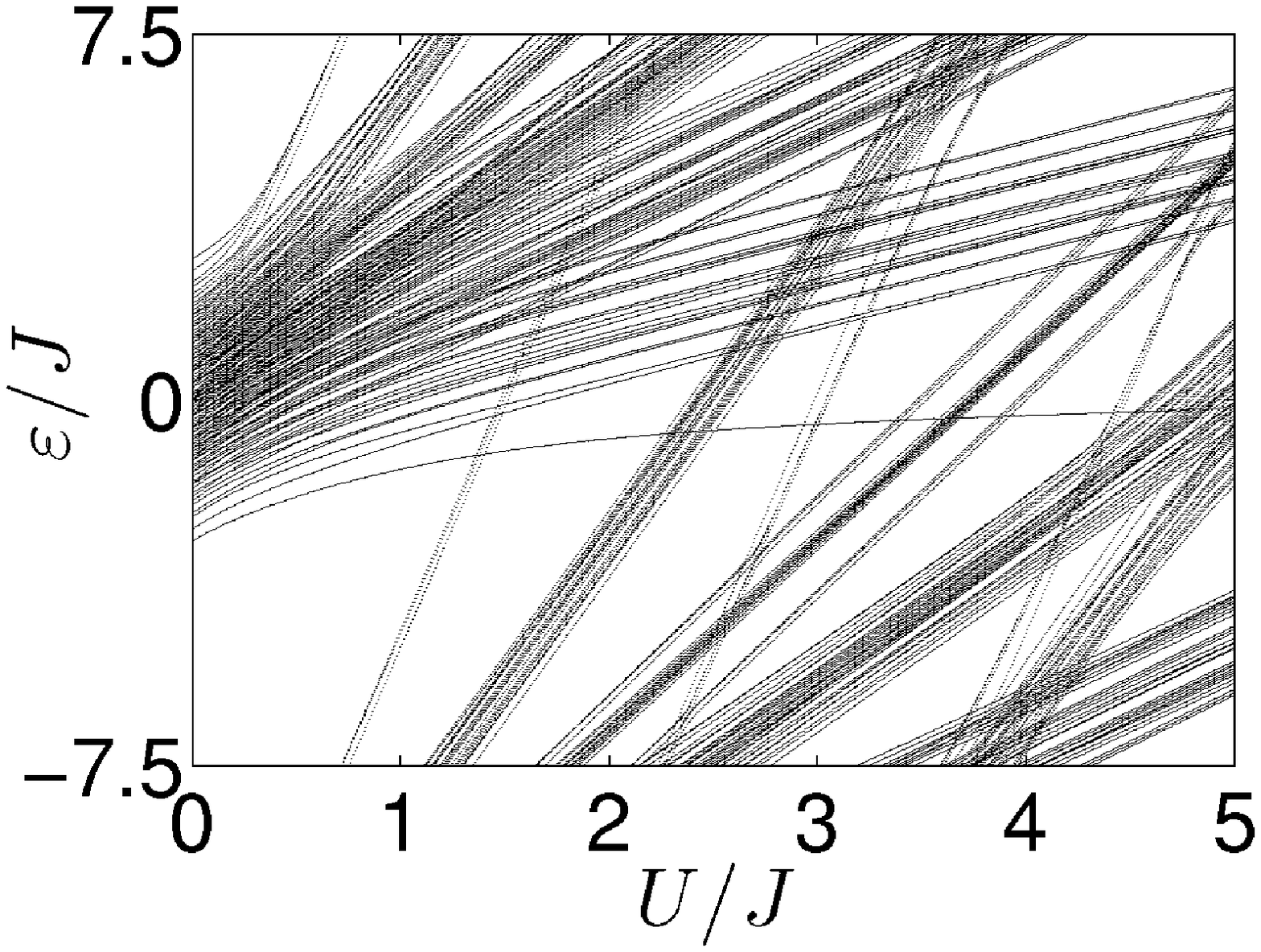}}
\caption{\label{F_4}
	Exact quasienergy spectra versus interaction strength~$U/J$ 
	for a small system ($N=M=5$) for $\hbar\omega/J=15$, and 
	$K_0=\nu\hbar\omega$, with $\nu = 0,1,2$. Amplitudes 
	$K_\omega/\hbar\omega$ are chosen such that $|J_{\rm eff}| = J/3$. 
	Observe that these spectra closely resemble the energy eigenvalues 
	in Fig.~\ref{F_1}, when the axes in that figure are rescaled by $1/3$ 
	and the eigenvalues are taken modulo $\hbar\omega$.} 
\end{figure*}

To substantiate our reasoning, the lower row of Fig.~\ref{F_2} shows exact
quasienergy spectra, obained from numerical solutions of the eigenvalue
equation~(\ref{eq:EIG}) with $N = M = 5$, for the same cases as considered in 
the dynamical simulations displayed in the upper row of that figure. These 
spectra are well matched to the effective Hamiltonian~(\ref{eq:HEF}) with 
$\Delta K_0 = 0$; in particular, the observed band collapses are accurately
located at the zeros of the respective Bessel function ${\rm J}_\nu$. In 
the vicinity of these points the quasienergy of the effective ground state 
separates markedly from the bands of excited states, thus indicating the 
presence of a Mott-like regime. The width of the quasienergy bands reflects 
the strength of the effective tunneling contact: For $\nu = 0$, there is no 
tilt and the contact provided by $\hat{H}_{\rm tun}$ is active for small 
$K_\omega/\hbar\omega$, resulting in wide bands. In contrast, for $\nu = 1$ 
and $\nu = 2$ tunneling is hindered by the static tilt, so that superfluidity 
is enabled by a process akin to photon-assisted tunneling only when the 
AC drive is sufficiently strong.

Moreover, $\hat{H}_{\rm eff}$ implies a simple scaling property: Quasienergy 
spectra obtained for different $\nu$ and different $K_\omega/\hbar\omega$, 
but corresponding to the same value of $|J_{\rm eff}|$, should be almost 
identical. This fact is confirmed in Fig.~\ref{F_4}, where we have plotted 
spectra for $\nu = 0$, $1$, and $2$, and driving amplitudes
$K_\omega/\hbar\omega = 1.811$, $0.711$ and $1.915$, respectively, chosen
such that $|J_{\rm eff}| = J/3$ in each case. Indeed, the three spectra
look strikingly alike; as expected; their gross features are obtained from
the energy spectrum shown in Fig.~\ref{F_1} by rescaling the axes of that
figure by a factor of $1/3$, and then taking the eigenvalues modulo
$\hbar\omega$ to account for the quasienergy Brillouin zones. 

Besides the existence of many-body Floquet states, a second key ingredient
to our proposal is adiabatic following. In fact, Floquet states respect an
adiabatic principle~\cite{BreuerHolthaus89}; the initial energy eigenstate 
evolves into the ``connected'' Floquet state when the AC drive is turned 
on sufficiently slowly. However, in the present scenario adiabatic following 
is endangered by the couplings associated with $\hat{V}$, which we have 
neglected when retaining only the diagonal blocks of the quasienergy 
matrix~(\ref{eq:QEM}). These couplings lead to many weak resonances among 
the instantaneous quasienergy levels. Hence, what appears as effectively 
adiabatic motion actually should be viewed as highly intricate Landau-Zener 
dynamics at multiple avoided level crossings, so that one may even deteriorate 
the quality of adiabatic following by reducing the rate of the parameter 
change in some cases. In the examples shown in Fig.~\ref{F_2}, the squared
overlap $ | \langle \psi_{\rm eff} | \psi(t) \rangle |^2 $ of the true wave 
function $|\psi(t)\rangle$ with the ground state $|\psi_{\rm eff}\rangle$
of the instantaneous $\hat{H}_{\rm eff}$ (with $\Delta K_0 = 0$) at
integer $t/T$ always remains larger than $0.26 \approx 0.89^N$, 
$0.64 \approx 0.95^N$, and $0.30 \approx 0.87^N$ for $\nu = 0$, $1$, and $2$, 
respectively, and $N = 9$. The extent to which this adiabatic principle can 
be exploited to guide the evolution of ultracold atoms in AC-driven optical 
lattices into certain desired target states is a subject requiring further 
studies, both experimental and  theoretical.

\section{Experimental feasibility and conclusion}
In a laboratory experiment with optical lattices, some restrictions must be 
met in order not to leave the scope of the single-band Bose--Hubbard model: 
The static tilt must remain weak enough to prevent Zener tunneling on the 
time scale of the experiment, and both the frequency and the amplitude of 
the AC force must remain sufficiently low in order not to excite transitions
to higher bands. Denoting the energy gap between the lowest two Bloch bands 
by $E_{\rm g}$, this translates into the requirements 
$U, J \ll K_0, \hbar\omega, K_\omega \ll E_{\rm g}$, while $U/J < (U/J)_c$. 
Considering a one-dimensional lattice with a depth of four atomic recoil 
energies and strong transversal confinement, as in the Z\"urich
experiment~\cite{StoeferleEtAl04}, one finds $E_{\rm g} \approx 23 \, J$
and $U/J \approx 614 \, a/d$, where $a$ denotes the $s$-wave scattering
length of the atomic species employed, and $d$ is the lattice constant.
Thus, favourable conditions with $U/J \lesssim 1$ are obtained when
$a/d \lesssim 0.0016$. This demand appears severe, but not impossible to 
satisfy.  

In closing, we remark that the modification~(\ref{eq:JEF}) of the hopping 
matrix element is very similar to the modification of atomic $g$-factors 
by oscillating magnetic fields~\cite{HarocheEtAl70}, both mathematically 
and conceptually~\cite{HolthausHone96}. Understanding that modificaton 
has been crucial for developing the ``dressed atom''-picture of atomic 
physics. Hence, the pioneering experiment~\cite{LignierEtAl07}, having
confirmed eq.~(\ref{eq:JEF}) for $\nu = 0$ with a Bose--Einstein condensate, 
eventually might lead to an analogous picture of ``dressed matter waves''.  

This work was supported in part by the Deutsche Forschungsgemeinschaft 
through the Priority Programme SPP~1116. We are grateful to Dr. Oliver Morsch
for communicating details of his experiment~\cite{LignierEtAl07} prior to 
publication. A.E.\ acknowledges a fellowship from the Studien\-stiftung des 
deutschen Volkes.

\end{document}